\documentclass[aps,prb,twocolumn,showpacs,float,nofootinbib,reprint,superscriptaddress]{revtex4-1}

\usepackage{amsmath}
\usepackage[mathscr]{euscript}
\usepackage{gensymb,amsfonts}
\usepackage{amssymb}
\usepackage{array}
\usepackage[T1]{fontenc}
\usepackage{lmodern}
\usepackage{tabularx}
\usepackage{fancyhdr}
\usepackage{graphicx}
\usepackage[english]{babel}
\usepackage{amsthm,mathtools}
\usepackage{braket}
\usepackage{tikz}
\usepackage{amsbsy}
\usepackage{mathrsfs}
\usepackage{epsfig}
\usepackage{array}
\usepackage{textcomp}
\usepackage{color}
\usepackage{bm,hyperref}
\usepackage[justification=raggedright,font=small]{caption}
\usepackage[titletoc,toc,title]{appendix}
\usepackage{ulem}
\usepackage{dcolumn}
\usepackage{ifpdf}
\usepackage{float}
\usepackage[font=small]{subcaption}
\usepackage{color}
\usepackage{framed}

\newcommand{\mP}{\mathcal{P}}
\newcommand{\beq}{\begin{equation}}
\newcommand{\eeq}{\end{equation}}

\graphicspath{{images/}}

\begin{document}

\title{Topological States on Fractal Lattices}

\author{Shriya Pai} 
\thanks{shra0375@colorado.edu}
\affiliation{
Department of Physics and Center for Theory of Quantum Matter,
University of Colorado, Boulder, CO 80309, USA
}

\author{Abhinav Prem}
\thanks{aprem@princeton.edu}
\affiliation{
Princeton Center for Theoretical Science,
Princeton University, NJ 08544, USA
}

\begin{abstract}
We investigate the fate of topological states on fractal lattices. Focusing on a spinless chiral $p$-wave paired superconductor, we find that this model supports two qualitatively distinct phases when defined on a Sierpinski gasket. While the trivial phase is characterized by a self-similar spectrum with infinitely many gaps and extended eigenstates, the novel ``topological'' phase has a \textit{gapless} spectrum and hosts chiral states propagating along edges of the graph. Besides employing theoretical probes such as the real-space Chern number, inverse participation ratio, and energy-level statistics in the presence of disorder, we develop a simple physical picture capturing the essential features of the model on the gasket. Extending this picture to other fractal lattices and topological states, we show that the $p+ip$ state admits a \textit{gapped} topological phase on the Sierpinski carpet and that a higher-order topological insulator placed on this lattice hosts gapless modes localized on corners. 
\end{abstract}

\maketitle

\section{Introduction}
\label{sec:intro}

The discovery of electronic insulators with topologically nontrivial band structures has led to remarkable progress in understanding gapped quantum phases. The prediction and experimental discovery of topological insulators (TIs)~\cite{kanemele,bernevigzhang,fukanemele,moorebalents,roy2009,konig2007,hsieh2008} and topological superconductors (TSCs)~\cite{readgreen,ivanov,stoneroy,zhang2018} led to a classification of gapped phases of non-interacting fermions~\cite{ryu2010,kitaev2009}; this ten-fold way encodes whether a system may host topologically nontrivial phases given the spatial dimension and the symmetries under which it is invariant. The nontrivial band topology of electronic states is manifest in striking universal properties, including robust gapless modes confined to the sample boundary and quantized response coefficients~\cite{hasankanermp,qizhangrmp}.

These concepts were later extended to crystalline symmetries, such as reflection, inversion, or rotation. Gapped phases protected by these symmetries are called topological crystalline insulators (TCIs)~\cite{fuTCI,hsiehTCI,okada2013,sessi2016,ma2017} and include higher-order topological insulators (HOTIs)~\cite{schindlerHOTI,benalcazar2017,langbehnHOTI,songHOTI,khalaf2018}. Specifically, an $n^{th}$ order TI/TSC in $d$ spatial dimensions is gapped everywhere except on a $d-n$ dimensional surface. More generally, TI/TSCs and HOTIs are examples of symmetry-protected topological (SPT)~\cite{chen2013,guwen2014} and crystalline SPT (cSPT)~\cite{song2017,buildingblock,elsethorngren} phases respectively, whose classification also accounts for interactions. Such phases have a trivial gapped bulk but host boundary (or hinge/corner) modes protected against local, symmetry-preserving perturbations~\cite{senthilSPT}.

A defining feature of topological phases is their robustness against disorder: provided the spectral (or mobility) gap remains finite and the disorder respects the symmetry protecting the TI/TSC, quantized coefficients and gapless edge modes persist~\cite{mong2012,ringel2012,schubert,foster2014}. Despite disorder breaking the lattice symmetries protecting TCIs, their boundary modes can evade localization when the full ensemble of disorder configurations remains symmetric~\cite{fulga2014,prodan2015}. Traditionally, robustness of topological states is established by adding disorder to a clean system, thereby assuming an underlying periodic reference state. This approach, while efficacious, fails when no such structure exists \textit{i.e.,} for aperiodic systems, including amorphous, quasiperiodic, and fractal systems. Nonetheless, topological phenomena have been shown to exist in both amorphous~\cite{agarwala2017,shibaglass,mitchell2018,bourne2018,agarwala2019,ojanen2019} and quasiperiodic~\cite{kraus1,kraus2,loring2016,bandres2016,quasicrystalQSH,spinbott} systems. 
\begin{figure}[b]
\centering
\begin{subfigure}[t]{0.23\textwidth}
\includegraphics[width=\textwidth]{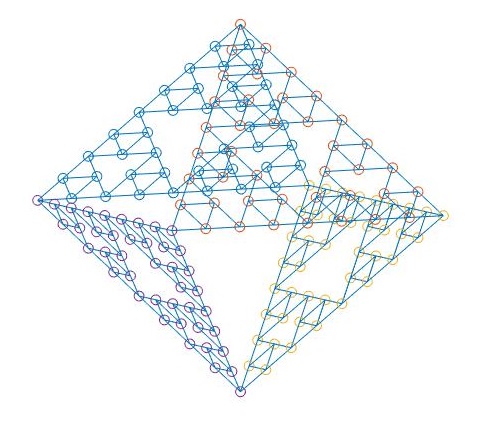}
\caption{}
\label{fig:gasket}
\end{subfigure}\quad
\begin{subfigure}[t]{0.23\textwidth}
\includegraphics[width=\textwidth]{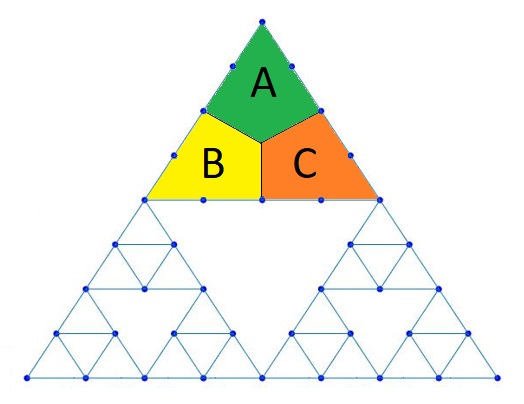}
\caption{}
\label{fig:partition}
\end{subfigure}
\caption{(a) SG with ``periodic'' boundary conditions, such that all sites have coordination number four. (b) Regions $A,B,C$ considered in real-space Chern number calculations.}
\end{figure}

That the topology of quantum states can be defined in the absence of spatial regularity over long distances opens the door to finding topological phases on fractal lattices, which lack a natural distinction between bulk and boundary, and whose (typically non-integer) Hausdorff dimensions differ from their topological dimensions. Interest in fractal structures, which have a rich history~\cite{recursion,banavar,gefen1,gefen2,rammal,alexander}, has been revived given experimental advances in creating and manipulating synthetic lattices with arbitrary structures, in both photonic and electronic systems \cite{kollarhyper,lieblattice,pband,collins2017,girovsky,drost}. In particular, fractal lattices have been fabricated using focused ion beam milling~\cite{herek}, molecular chains~\cite{shang,tait2015,zhang2016}, and scanning-tunneling-microscopy (STM) techniques~\cite{kempkes2018}, with theoretical studies primarily focusing on localization and transport phenomena~\cite{vanVeen1,vanVeen2,pal2012,kosior2017,bpal2018,genzor2019}.
 
 \begin{figure*}[t]
 \centering
 \includegraphics[width=0.8\textwidth]{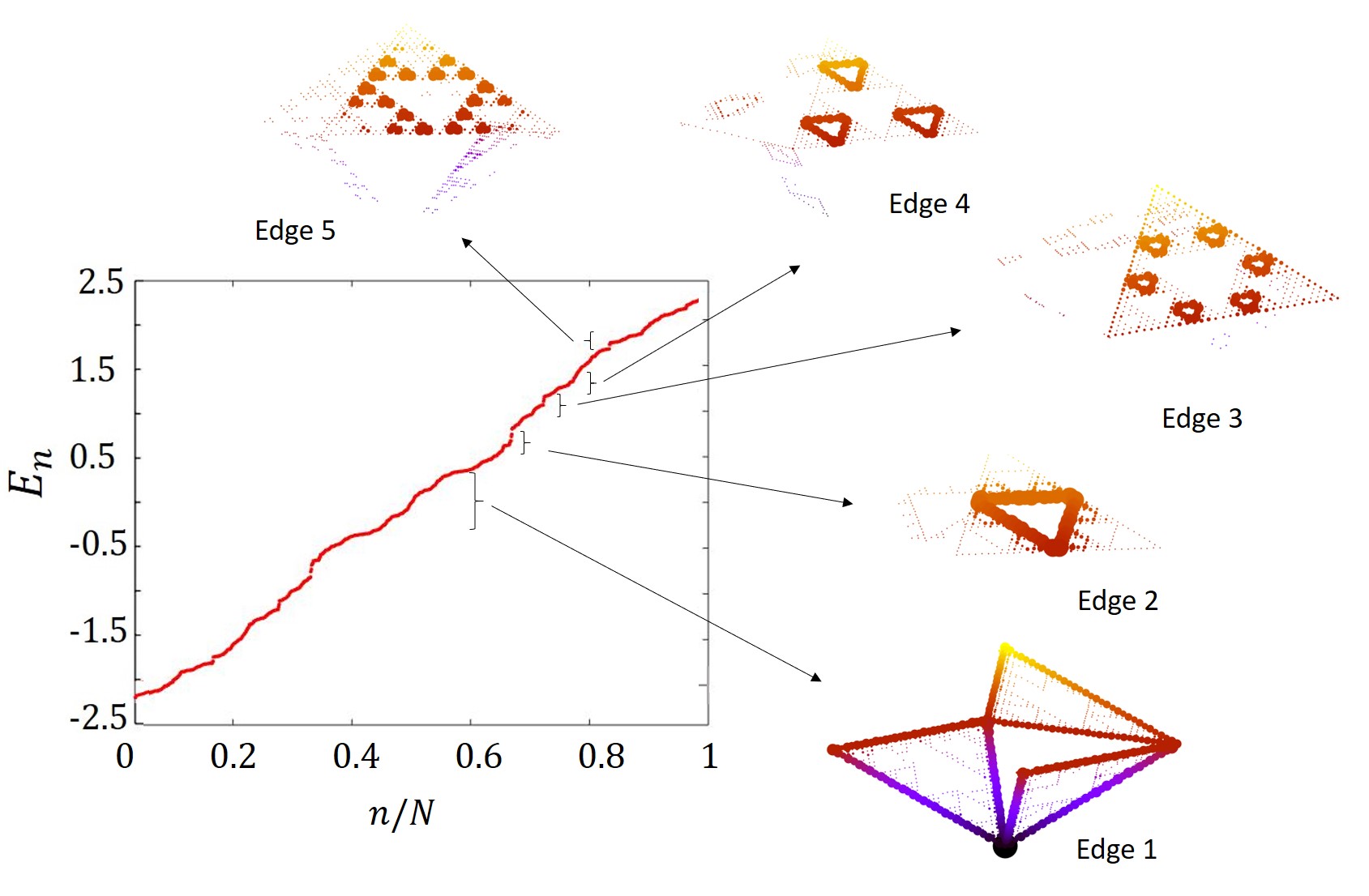}
 \caption{Gapless topological phase of the chiral $p+ip$ superconductor on the SG, with $g=5,\Delta=1,t=0.5, \mu=0.5$. Energy spectrum and probability densities of eigenvectors at indicated energies are shown. Color scale indicates values of $x,y,z$ coordinates, with dot size indicating the magnitude of the probability density at that point.}
 \label{fig:spectrum}
 \end{figure*}
 
However, our understanding of the influence of self-similar geometry on the topological character of electronic states remains nascent, having received attention only recently~\cite{fractalized,neupert}. In this paper, we fill this lacuna by developing a general framework elucidating the fate of topological states on fractal lattices embedded in two dimensions (2D). Through this picture, we find that the nature of thermodynamic phases---gapped vs gapless---on fractal lattices depends crucially on the ratio of bulk to edge coordinated sites. Focusing on the chiral $p$-wave superconductor on the Sierpinski gasket, we show that qualitative features obtained through numerical diagonalization can be understood simply through our framework. Besides characterizing the two distinct phases of this model using various theoretical tools, we further corroborate our understanding by studying both the $p$-wave superconductor and an HOTI on the Sierpinski carpet. 

\section{Model}
\label{sec:model}

\begin{figure*}[t]
 \centering
 \includegraphics[width=0.8\textwidth]{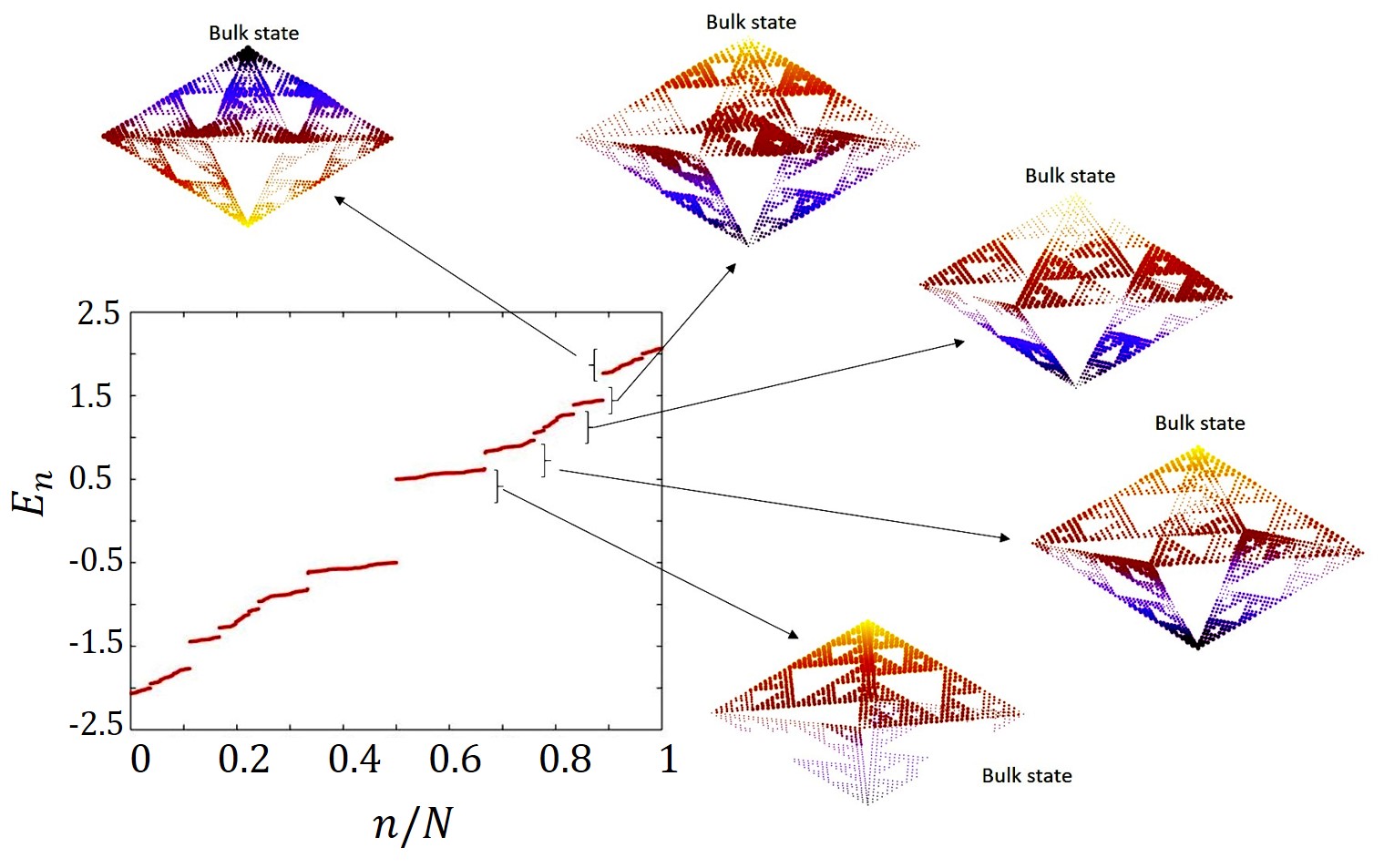}
 \caption{Gapped trivial phase of the $p+ip$ superconductor on the SG, with $g=5,\Delta=1,t=0.5, \mu=2$. Energy spectrum and probability densities of eigenvectors at indicated energies are shown. Color scale indicates values of $x,y,z$ coordinates, with dot size indicating the magnitude of the probability density at that point.}
 \label{fig:spectrumTrivial}
 \end{figure*}

We consider a 2D spinless chiral $p$-wave superconductor (symmetry class D~\cite{ryu2010}) within the Bogoliubov-deGennes (BdG) framework, with the mean field lattice BCS Hamiltonian:
\beq
\label{bdgHam}
\hat{H} = -t \sum_{\langle \bf{r}, \bf{r'} \rangle} \hat{c}_{\bf{r}}^\dagger \hat{c}_{\bf{r'}} - \mu \sum_{\bf{r}}  \hat{c}_{\bf{r}}^\dagger \hat{c}_{\bf{r}} + \sum_{\bm{r},m} [ \Delta_m \hat{c}_{\bf{r} + \bm{e}_m}^\dagger \hat{c}_{\bf{r}}^\dagger + \text{h.c.} ], 
\eeq
where $\hat{c}_{\bf{r}}^\dagger,\hat{c}_{\bf{r}}$ satisfy fermionic anti-commutation relations $\{\hat{c}_{\bf{r}},\hat{c}_{\bf{r'}}^\dagger \} = \delta_{\bf{r},\bf{r'}}$, $t$ is the nearest-neighbor hopping, $\mu$
is the chemical potential, and we set the lattice spacing $a = 1$. Specifying to a triangular lattice~\cite{cheng2010}, the pairing term $\Delta_m = \Delta e^{i \pi m/3}$ is defined on the nearest-neighbor bonds corresponding to the three lattice vectors $\bf{e}_m$ with azimuthal angles $m \pi/3$ ($m=0,1,2$). We introduce the standard Bogoliubov transformation: $\hat{c}_{\bf{r}} = \sum_{\bf{r}} \left[u_{n,\bf{r}} \hat{\gamma}_n + v_{n,\bf{r}} \hat{\gamma}_n^\dagger \right]$, where $\hat{\gamma}_n$ is the Bogoliubov quasiparticle annihilation operator and $(u_{n,\bf{r}}, v_{n,\bf{r}})^T$ diagonalizes the BdG Hamiltonian~\eqref{bdgHam}, with eigenvalue $E_n$.

We study this model on a Sierpinski gasket (SG) with ``periodic'' boundary conditions (see Fig.~\ref{fig:gasket}) \textit{i.e.,} with four gaskets arranged on alternating faces of an octahedron, ensuring that all lattice sites are equally (four) coordinated. We construct a lattice regulated (with a smallest triangle) SG recursively, by adding sites/bonds to a gasket at generation $g$ to arrive at the $g+1$ SG. The largest lattice we can probe numerically has $g=6$, with the total number of sites $N \sim 3^{g+1}$ at generation $g$.
 
Setting $\Delta > 0$ and noting that the Hamiltonian Eq.~\eqref{bdgHam} admits a topological phase on a triangular lattice for $-6t < \mu < 2t$ (see Appendix \ref{app:triangle}), we find that this model admits topologically distinct phases even on the SG. The qualitative distinction between the two phases is illustrated in Fig.~\ref{fig:spectrum}, which shows the spectrum and states for $g = 5$. For $\mu>2t$ or $\mu<-6t$, we find a fully gapped ``trivial'' phase (see Fig. \ref{fig:spectrumTrivial}), where eigenstates are delocalized, thereby behaving as bulk states in ordinary gapped systems. In the thermodynamic ($g \to \infty$) limit, the spectrum is self-similar, with infinitely many gaps. In contrast, for $-6t<\mu<2t$ we find that the amplitude of the largest gap in the spectrum decays exponentially with increasing generation (see Appendix \ref{gap:scaling}), such that the spectrum is strictly \textit{gapless} in the $g\to \infty$ limit. Thus, this parameter range describes a qualitatively distinct phase with emergent \textit{continuous} scale invariance, unlike the trivial phase which only possesses \textit{discrete} scale invariance. Particle-hole symmetry is present in both phases. While the spectra are obtained by numerically diagonalizing the BdG Hamiltonian~\eqref{bdgHam}, these can in principle also be obtained recursively (see Appendix \ref{app:recursive} for details).

An intriguing feature of the gapless phase is the \textit{edge-like} nature of eigenstates: in Fig.~\ref{fig:spectrum}, we plot the electronic densities for representative states at the indicated energies, revealing states sharply localized on triangular motifs formed by sites of various generations \textit{i.e.,} localized around the inner edges (or holes) of the SG. While states closest to $E = 0$ are localized on the outer edges, corresponding to the earliest generations, there is a hierarchy of states localized on inner edges created at subsequent generations of the SG. In the thermodynamic limit, we expect that \textit{all} eigenstates in this phase will be sharply localized along edges. Remarkably, these localized states are also \textit{chiral}, with a wave-packet initialized on any inner edge propagating in the direction opposite to that of one initialized on the outermost edge (see Appendix \ref{app:chiral}). 

Surprisingly, we find that the transition between the trivially gapped and the gapless phase coincides with the trivial $\leftrightarrow$ topological transition of Eq.~\eqref{bdgHam} on the triangular lattice. This observation hints that the model on the SG inherits its behavior from one defined on a triangular lattice. Indeed, we can regard the inner edges of the SG as holes in a triangular lattice, which, in the topological phase of Eq.~\eqref{bdgHam}, host gapless chiral Majorana modes propagating counter to the outermost edge state~\cite{aliceareview}. Since the number of these holes increases with $g$, there are infinitely many gapless modes in the spectrum as $g\to \infty$, resulting in a gapless spectrum. This physical picture suggests that the chiral eigenstates in the gapless phase are descended from Majorana edge modes of the $p+ip$ state on a triangular lattice. We hence dub this the gapless \textit{topological} phase on the SG.

\begin{figure*}[t]
\centering
\begin{subfigure}[t]{0.45\textwidth}
\includegraphics[width=\textwidth]{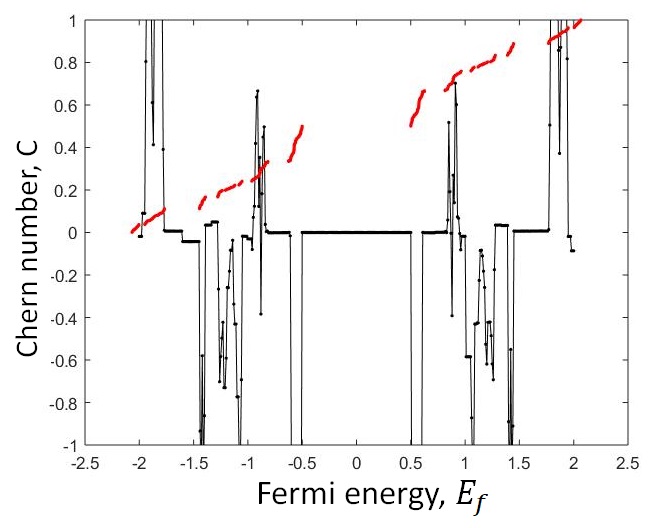}
\caption{}
\label{fig:cherntriv}
\end{subfigure}\quad
\begin{subfigure}[t]{0.45\textwidth}
\includegraphics[width=\textwidth]{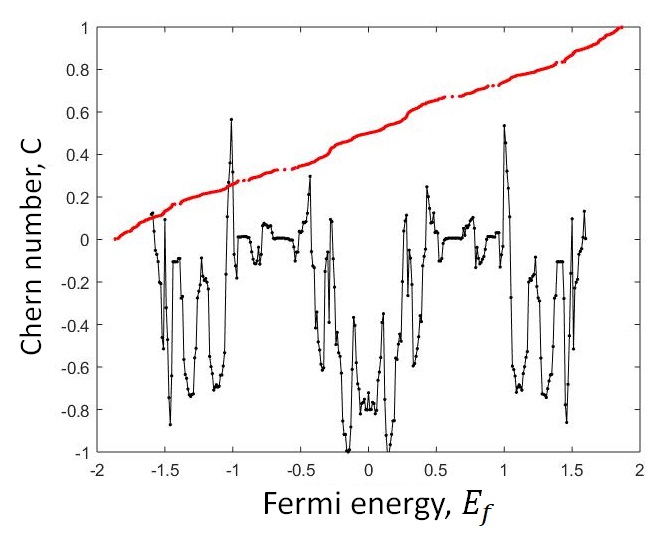}
\caption{}
\label{fig:chernTop}
\end{subfigure}
\caption{The real-space Chern number (black curve) as a function of Fermi energy $E_{f}$, with the corresponding spectrum shown in red in (a) the trivial phase ($\mu=2$), and (b) the topological phase ($\mu=0.5$). Here, $g=4$, $t=0.5$, and $\Delta=0.5$. }
\end{figure*}

\section{Diagnostics}
\label{sec:diagnostics}

Before building on this intuitive picture and showing that it generalizes to other fractal lattices, such as the Sierpinski carpet (SC), and other topological states, we further characterize the two distinct phases of the $p+ip$ superconductor on the SG using some standard diagnostics.

\subsection{Real-space Chern number} 
\label{ssec:realSpaceChernNo}

Since our model lacks translation invariance, and only retains (discrete) scale invariance, we cannot use the momentum-space Chern number to characterize the topological and trivial phases of the $p+ip$ superconductor on the SG. Thus, we instead compute the real-space Chern number introduced in Ref.~[\onlinecite{kitaev2006}], which reduces to the momentum-space Chern number in the presence of translation invariance:
\begin{equation}
    \mathcal{C}= 12 \pi i \sum_{j\in A}\,\sum_{k\in B}\,\sum_{l\in C}(\mP_{jk}\mP_{kl}\mP_{lj} - \mP_{jl}\mP_{lk}\mP_{kj}),
\label{realspace}
\end{equation}
where $\mP$ projects onto occupied states with respect to a given chemical potential, and $j,k,l$ are indices corresponding to three distinct neighboring regions $A,B,C$, arranged counter-clockwise (see Fig.~\ref{fig:partition}). In Eq. \ref{realspace}, $\mathcal{P}_{ij}$ is a $2\times 2$ matrix whose rows correspond to $c^{\dagger}_{i}, c_{i}$, and whose columns correspond to  $c_{j}, c^{\dagger}_{j}$. Retaining the site basis, we rotate only the $k$ /pseudospin basis. We then diagonalize the $2 \times 2$ matrix in the expression for $\mathcal{C}$ such that pseudospin is now a good quantum number, and then take the trace. With $\widetilde{\mathcal{P}}_{ij}$ representing the $2 \times 2$ block after diagonalization, the expression for the Chern number can be rewritten as:

\begin{equation}
    \mathcal{C}= 12 \pi i \sum_{j\in A}\,\sum_{k\in B}\,\sum_{l\in C}\textrm{Tr}(\widetilde{\mathcal{P}}_{jk}\widetilde{\mathcal{P}}_{kl}\widetilde{\mathcal{P}}_{lj} - \widetilde{\mathcal{P}}_{jl}\widetilde{\mathcal{P}}_{lk}\widetilde{\mathcal{P}}_{kj}),
\label{chernTrace}
\end{equation}

For $g=5$ in the trivial phase, we find that $\mathcal{C} = 0$ for all gapped regions of the spectrum (see Fig. \ref{fig:cherntriv}). We have checked that this quantization becomes independent of the specific choice of regions $A,B,C$ at large $g \geq 4$ \textit{i.e.,} in the limit when the number of sites in each region becomes large. In the thermodynamic ($g \to \infty$) limit, the spectrum within the trivial phase displays an infinite hierarchy of self-similar gaps, and we expect that $\mathcal{C}$ will vanish identically for each of the infinitely many gaps in the spectrum.

In contrast, within the topological phase the gapped regions of the SG scale to zero and have a trivially quantized Chern number. As can be seen in Fig.~\ref{fig:chernTop}, we find that indeed $\mathcal{C}=0$ within the finite-size gaps at finite $g$ in the topological phase. Nevertheless, similarly to previous works on topological amorphous superconductors~\cite{ojanen2019} and on the quantum Hall effect on fractal lattices~\cite{neupert}, we expect the Chern number to take a non-trivial quantized value within the gapless regions due to the presence of a mobility gap and the topological nature of the phase. While our numerics suggest that the Chern number tends towards a quantized non-zero value with increasing $g$ in regions corresponding to low but non-zero density of states, we are numerically limited to $g \leq 5$, for which finite-size effects obscure the expected quantization.

Thus, in the thermodynamic limit, the trivial phase will exhibit a strictly quantized $\mathcal{C} = 0$ within the infinitely many gaps in the spectrum; on the other hand, although the spectrum becomes gapless in the topological phase, we expect that $\mathcal{C}$ converges to a non-trivial quantized value as $g \to \infty$ in the gapless regions due to the presence of a mobility gap~\cite{neupert}. Verifying the latter requires investigating the model on a SG with large $g$, which is beyond our current numerical capabilities.

\begin{figure*}[t]
\centering
\begin{subfigure}[t]{0.45\textwidth}
\includegraphics[width=\textwidth]{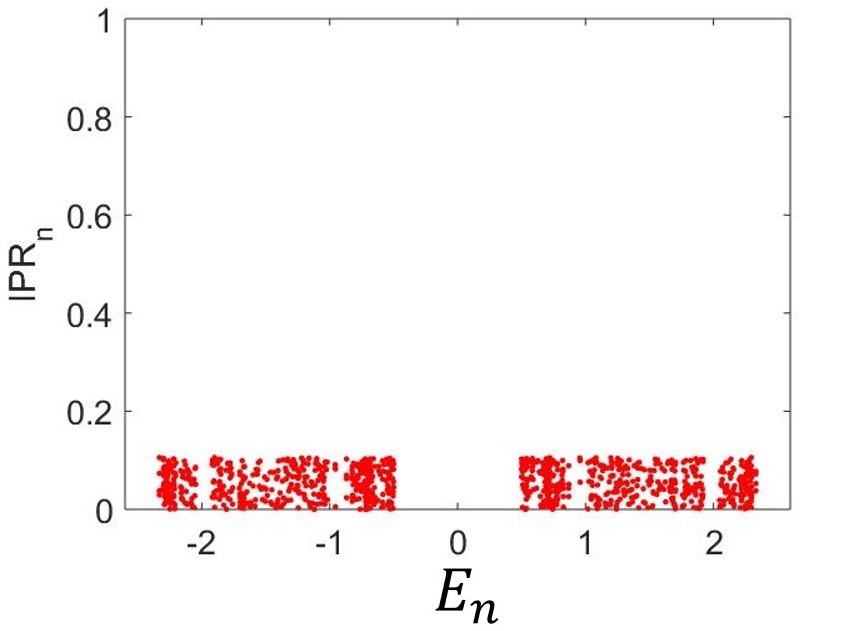}
\caption{}
\label{fig:iprtriv}
\end{subfigure}\quad
\begin{subfigure}[t]{0.45\textwidth}
\includegraphics[width=\textwidth]{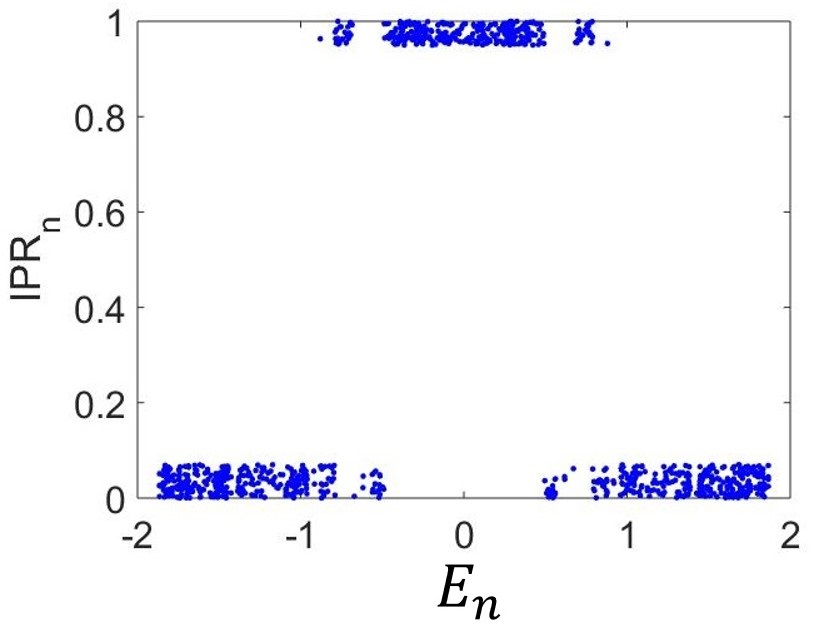}
\caption{}
\label{fig:iprtop}
\end{subfigure}
\caption{IPR$_n$ (using open boundary conditions on a single SG), with $g = 4, t = 0.5, \Delta = 0.5$. All states are delocalized in the (a) trivial phase ($\mu = 2$) while the (b) topological phase ($\mu = 0.5$) exhibits states localized around edges.}
\end{figure*}

\begin{figure}[b]
 \centering
 \includegraphics[width=0.4\textwidth]{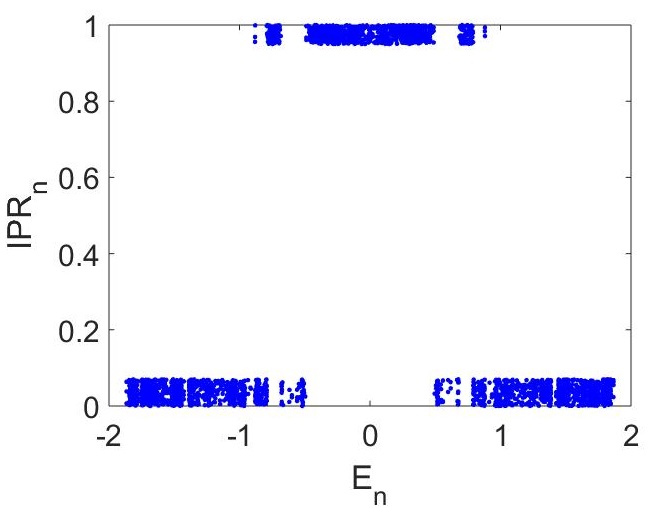}
 \caption{IPR$_{n}$ with $g=5, t=0.5, \Delta=0.5$. Comparison with Fig.~\ref{fig:iprtop} clearly shows that the number of localized states in the middle of the spectrum increases with $g$.}
 \label{fig:iprtopg5}
 \end{figure}

\subsection{Inverse participation ratio} 
\label{ssec:ipr}

\begin{figure}[b]
 \centering
 \includegraphics[width=0.4\textwidth]{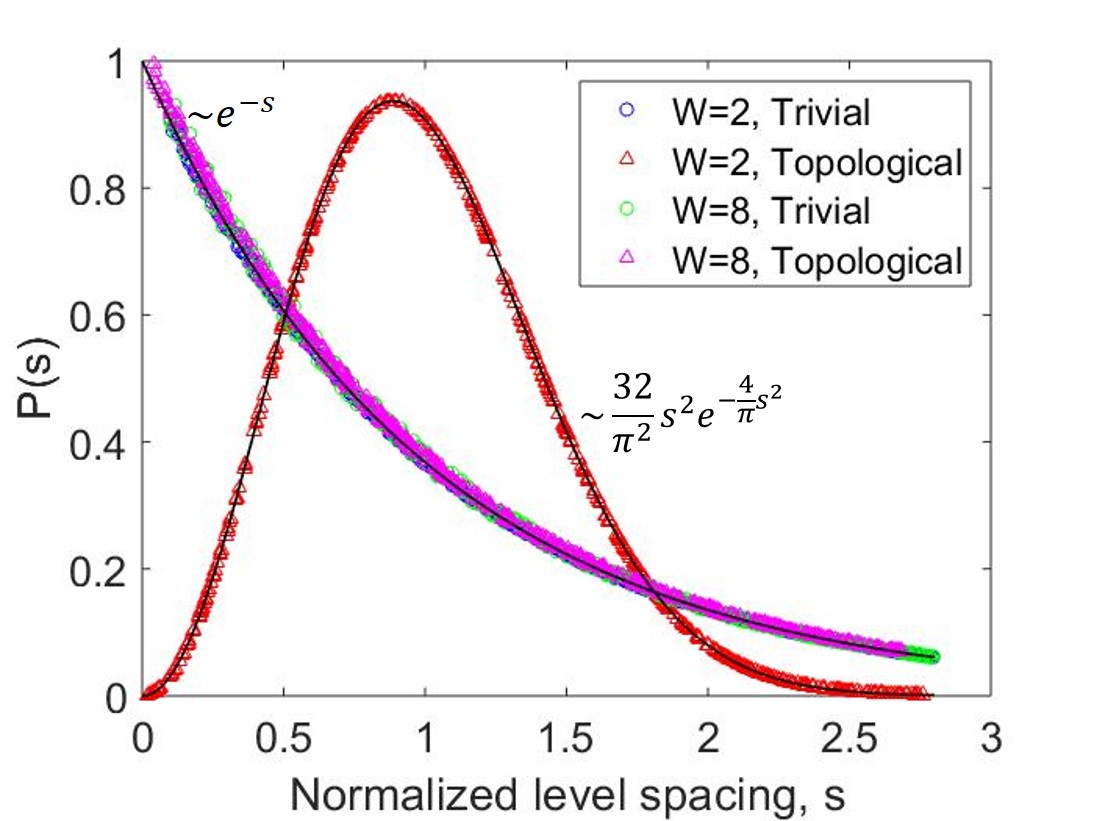}
 \caption{Distribution of normalized energy level spacings with disorder $W$, with $g = 4 , t = 0.5, \Delta = 0.5$. Level statistics shown for weak ($W=2$) and strong ($W=8$) disorder for the trivial ($\mu = 2$) and topological ($\mu = 0.5$) phase.}
 \label{fig:levelstatistics}
 \end{figure}
Another useful diagnostic is the inverse-participation-ratio (IPR) of the $n^{th}$ eigenstate~\cite{thouless1974,franz1996},
\beq
\textrm{IPR}_n = \frac{\sum_{\bf{r}} \left( |u_{n,\bf{r}}|^4 + |v_{n,\bf{r}}|^4 \right)  } { \big[ \sum_{\bf{r}}  \left( |u_{n,\bf{r}}|^2 + |v_{n,\bf{r}}|^2 \right) \big]^2}
\eeq
which scales as $L^{-2}$ for extended states but remains finite for localized states even in the thermodynamic limit. In the trivial phase, \textit{all} eigenstates are delocalized (see Fig.~\ref{fig:iprtriv}), reflecting their bulk nature. Increasing $g$ suppresses the IPR values further towards zero. In the topological phase, the IPR values instead abruptly jump between $\sim 0$ and $\sim 1$, with the latter corresponding to eigenstates localized along the various edges (or holes) of the SG, as in Fig.~\ref{fig:spectrum}. The number of localized states increases with $g$ (see Figs. \ref{fig:iprtop} and \ref{fig:iprtopg5}), consistent with the physical picture discussed above: cutting out holes from the triangular lattice does not introduce any edge modes in the trivial phase, and all states remain extended. In the topological phase however, additional gapless edge modes are introduced, with the number of such modes increasing with $g$. This agrees with the numerical observation of localized states with IPR$_n \sim 1$ as shown in Fig.~\ref{fig:iprtop}.

\subsection{Level Statistics}

Since disorder provides an independent probe of topology, we add an onsite term $\sum_{\bf{r}} V_{\bf{r}} \hat{c}_{\bf{r}}^\dagger \hat{c}_{\bf{r}}$ to the Hamiltonian~\eqref{bdgHam}, with $V_{\bf{r}}$ drawn randomly from the uniform distribution $[-W/2,W/2]$. In Fig.~\ref{fig:levelstatistics}, we plot the energy-level spacing distributions, averaged over 500 disorder realizations, for weak and strong disorder in both phases. The normalized level spacing is given by $s = |E_n - E_{n+1}|/\delta(E_n)$, with $\delta(E_n)$ the mean-level spacing near energy $E_n$. In the trivial phase, the distribution is Poissonian at both weak and strong disorder, consistent with a \textit{localized} phase. The level spacings in the topological phase follow unitary Wigner-Dyson (GUE) statistics at weak disorder ($W=2$) and transition to Poisson at strong ($W=8$) disorder, with the transition\footnote{Since we expect regions with quantized $\mathcal{C} \neq 0$ in the gapless topological phase (due to a mobility gap), the transition likely occurs through a critical delocalized state, as is the case for insulating systems with a quantized $\mathcal{C} \neq 0$ at small disorder~\cite{prodan2010}.} to the Anderson insulator occurring at $W \sim 5$. Agreement with the Wigner surmise (for $\beta =2$) at weak disorder (see Fig.~\ref{fig:levelstatistics}) indicates that the gapless topological phase is a \textit{diffusive metal}~\cite{mirlin2000,chou2014}.

\section{Recursive Decimation}
\label{sec:decimation}

\begin{figure}[b]
    \centering
    \includegraphics[width=0.5\textwidth]{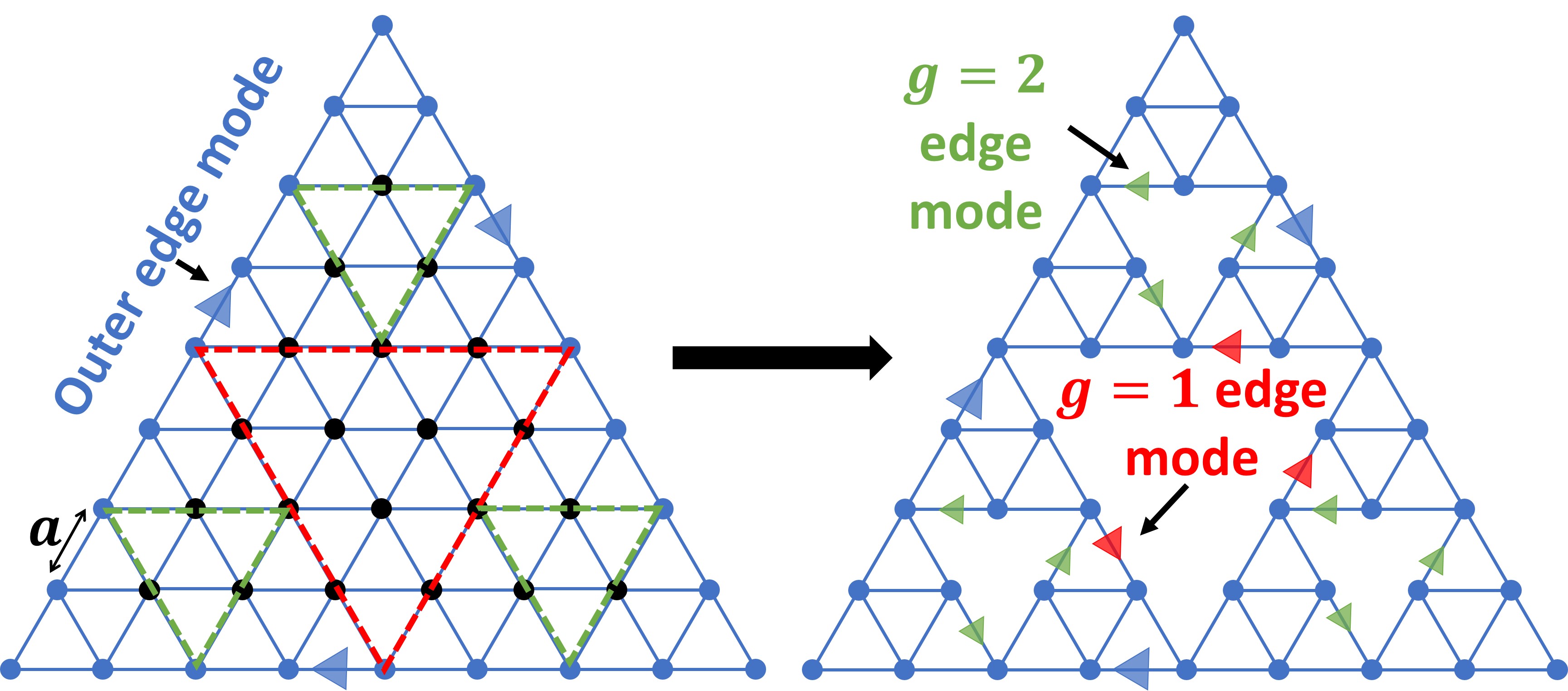}
    \caption{Decimating a triangular lattice recursively to generate the SG. Blue (black) dots denote sites with boundary (bulk) coordination. Sites and bonds inside the red (green) triangle(s) are eliminated at the first (second) step.}
    \label{fig:decimation}
\end{figure}

We propose a physical picture which elucidates how topological states on 2D fractal lattices inherit their behavior from a ``parent'' state on an underlying periodic lattice. Consider the BdG Hamiltonian~\eqref{bdgHam} on a triangular lattice with open boundary conditions, lattice spacing $a$, and size $L=2^p a$, as shown in Fig.~\ref{fig:decimation}. We define bulk and edge sites as those with coordination number six and four, respectively\footnote{Four copies of the lattice can be arranged as in Fig.~\ref{fig:gasket} such that the corner sites are also four coordinated.}. We now decimate sites and bonds recursively to generate the SG. At the $g^{th}$ step ($g\geq 1$), we eliminate all sites and bonds contained inside $3^{g-1}$ inverted triangles of length $L/2^g$, introducing an additional $3^{g-1}$ inner boundaries into the lattice. The procedure continues until $g=g_c$, with $L/2^{g_c} = 2a$ ($g_c = p-1$), at which stage a generation $g_c$ SG is produced: the ratio of bulk sites $n_B(g)$ to edge sites $n_E(g)$ vanishes identically when $g = g_c$ (see Appendix \ref{app:decimation} for details). This process is illustrated in Fig.~\ref{fig:decimation}. 

Starting in the topological phase, where a chiral Majorana mode propagates clockwise along the outermost boundary, each subsequent iteration introduces additional physical boundaries into the lattice, each hosting a chiral Majorana mode propagating counter-clockwise~\cite{aliceareview}. In the thermodynamic limit $L/a \to \infty$, the decimation is repeated infinitely many times ($g_c \to \infty$) until only boundary sites are left and a chiral Majorana mode propagates along each of the infinitely many edges, resulting in a gapless spectrum. Thus, the decimation picture shows that the chiral eigenstates of the gapless topological phase are intimately linked to the Majorana edge modes of the underlying $p$-wave state. Further, the absence of \textit{any} bulk sites explains why all bulk features of the underlying model are washed out as $g\to\infty$, with the novel gapless state effectively described by a self-similar network of chiral 1D Majorana modes.

\begin{figure*}[t]
\centering
\begin{subfigure}[t]{0.45\textwidth}
\includegraphics[width=\textwidth]{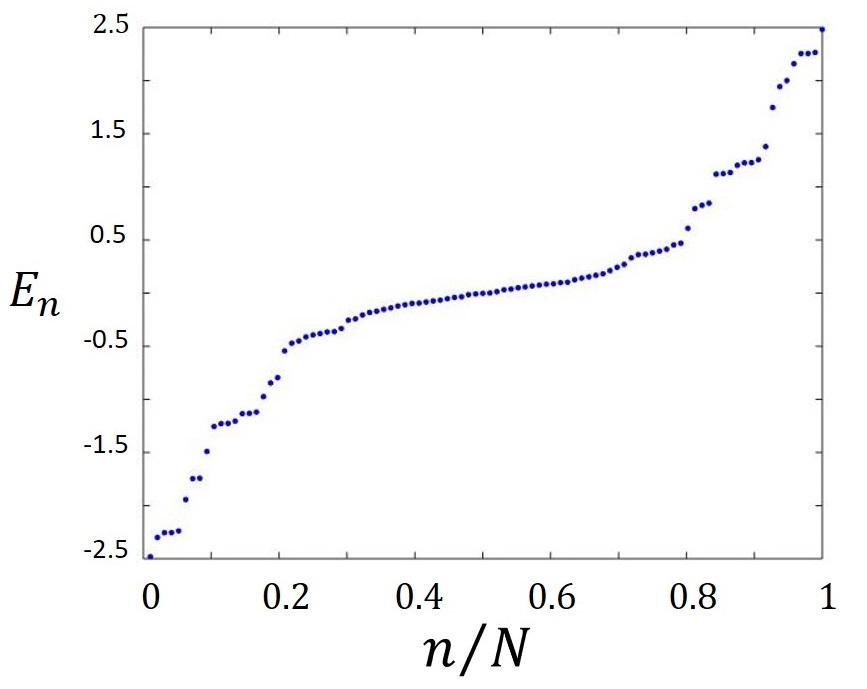}
\label{fig:hotiSpectrum}
\caption{}
\end{subfigure}\quad
\begin{subfigure}[t]{0.45\textwidth}
\includegraphics[width=0.8\textwidth]{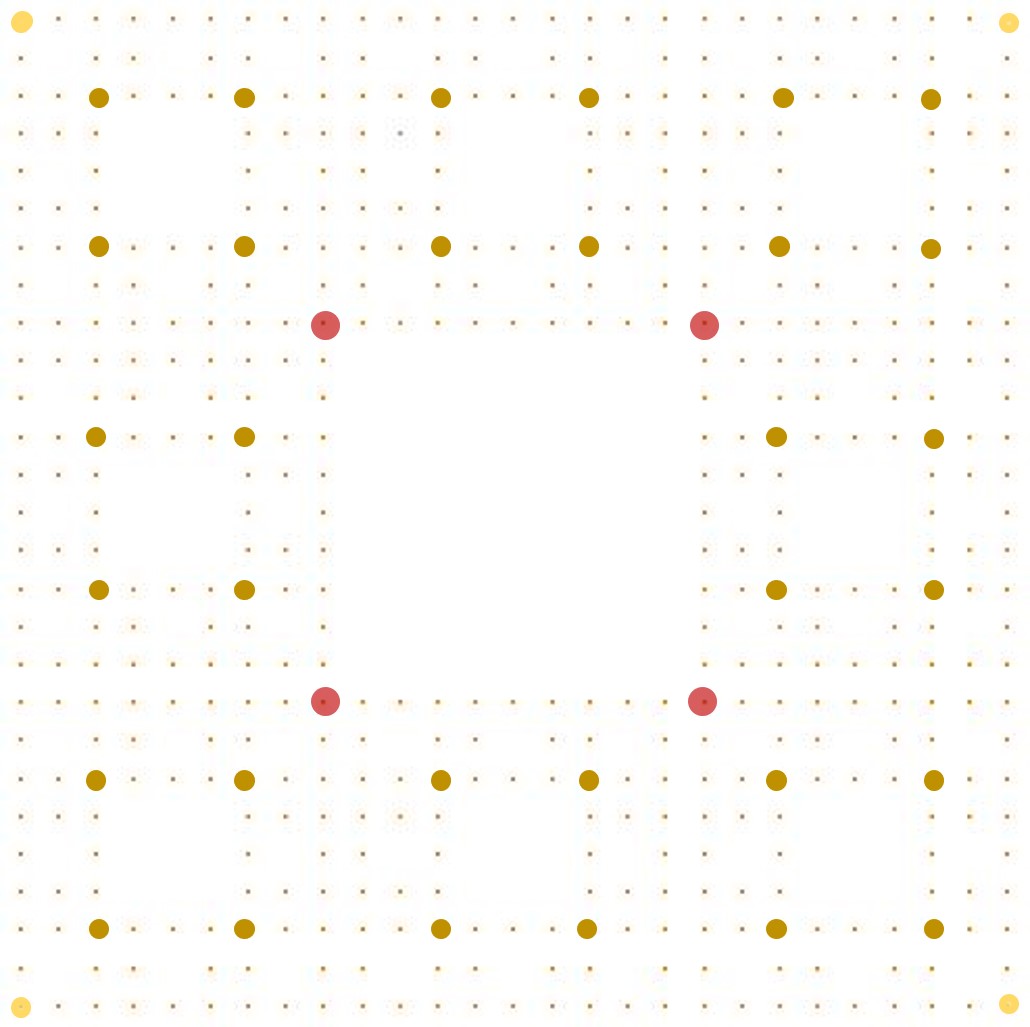}
\caption{}
\label{fig:hotiCornerMode}
\end{subfigure}
\caption{(a) Spectrum and (b) non-propagating corner modes in an HOTI defined on the SC ($g=3, \gamma=0.5, \lambda=1$).}
\label{fig:corner}
\end{figure*}

Starting instead in the trivial phase, each iteration only introduces additional gaps as no edge modes appear. The self-similar arrangement of the gaps is a consequence of discrete scale invariance of the generated SG, and the trivial $\to$ topological transition on the SG can be understood as the proliferation of chiral Majorana modes which occurs during the transition on the underlying periodic lattice. Our analytic picture naturally accounts for the phase boundaries of Eq.~\eqref{bdgHam} on the SG matching those on the triangular lattice. We also expect that the gapless topological phase inherits the robustness of the edge modes against arbitrary local perturbations respecting the symmetry protecting the parent ($p+ip$) state.

To further test our decimation picture, we place the $p+ip$ Hamiltonian on the SC. This lattice can be constructed by recursively decimating a square lattice, on which a topological phase exists for $-t<\mu< t$. However, the ratio $ \lim_{g\to \infty} n_B(g)/n_E(g) \sim 5$ for the SC, resulting in more bulk than edge coordinated sites. Crucially, the distance between gapless edge modes appearing along inner boundaries at each step of the decimation process \textit{decreases} with each iteration, such that each Majorana edge mode on the recursively generated SC is separated from one with opposite chirality by $3a$. In the thermodynamic limit, these edges states back-scatter and hybridize, leading to a gapped spectrum; we thus expect that bulk features of the underlying state persist on the SC as $g \to \infty$ even in the topological phase. Results obtained by numerically diagonalizing the BdG Hamiltonian Eq.~\eqref{bdgHam} on the SC vindicate our prediction: we find a trivial ($\mathcal{C} = 0)$ and a \textit{gapped} topological (with quantized $\mathcal{C} = 1$) phase, with phase boundaries matching those of the model on the square lattice (see Appendix \ref{app:carpet} for details).

We posit that the above analysis readily generalizes to \textit{any} parent 2D topological state protected by internal symmetries: for parameters corresponding to the topological phase on a triangular lattice, the model will admit a gapless topological phase on the SG, whose physics is governed by that of the 1D gapless edge states of the parent state. On the SC, for parameters corresponding to the topological phase on the square lattice, the spectrum will remain gapped and exhibit a nontrivial quantized topological invariant. Thus, the nature of topological states on a given fractal lattice depends crucially on whether $\lim_{g\to\infty}n_B(g)/n_E(g)$ remains finite or vanishes, resulting in a gapped or gapless topological phase respectively. The results of Ref.~[\onlinecite{fractalized}], which studied the half-BHZ model~\cite{bernevigbook} on the SG and SC, are in excellent agreement with our conjecture and support the generality of our arguments. 

\section{HOTI on the Sierpinski carpet}

Extending the above ideas to topological states protected by spatial symmetries requires more care, since we must ensure that no symmetries protecting the underlying state are broken at any step of the recursive decimation, in order to stay within the same phase. For instance, for a cSPT protected by $C_4$ rotation, we can start from a square lattice and recursively generate the SC through decimation, resulting in a gapped topological phase in the thermodynamic limit. To demonstrate the applicability of our general framework to this case, we have studied the paradigmatic four-band model of an HOTI, introduced in Ref.~[\onlinecite{benalcazar2017}], on the SC. The real space Hamiltonian on the square lattice is given by:
\beq
\label{cornerModeHamiltonian}
H = - \sum_{m,n} \left[ \lambda^{(1)}_{m,n} \hat{c}_{m+1,n}^\dagger \hat{c}_{m,n} +  \lambda^{(2)}_{m,n} \hat{c}_{m,n+1}^\dagger \hat{c}_{m,n} + \textrm{h.c.} \right] \nonumber,
\eeq
where $\hat{c}_{m,n}^\dagger, \hat{c}_{m,n}$ are fermionic creation/annihilation operators for site $(m,n)$ of the square lattice, and where
\begin{equation*}
2 \lambda^{(1);(2)}_{m,n} = \lambda (1 + (-1)^{m;n})  + \gamma (1 - (-1)^{m;n}) \, .
\end{equation*}
This model preserves $C_{4}$ rotation, time-reversal, and charge-conjugation symmetries, and presents localized corner modes when $|\gamma/\lambda|<1$. Starting from the topological phase on the square lattice, we recursively decimate the lattice to generate the SC. Each iteration creates additional inner boundaries, each hosting protected gapless corner modes since no symmetries are broken at any stage. Following our general arguments, we expect a gapped topological phase on the SC as $g\to\infty$, with modes localized along the corners of infinitely many inner edges. As shown in Fig.~\ref{fig:corner}, we indeed find a gapped spectrum and corner modes on \textit{all} inner boundaries. While we are numerically limited to $g=3$, we expect this behavior persists for larger generations. We also note that the topological nature of the SC HOTI is protected only in the presence of a particle-hole symmetry in addition to a $C_4$ symmetry: in the absence of particle-hole symmetry, the zero energy modes can be shifted around without breaking the $C_4$ symmetry~\cite{benalcazar2017}. Besides the generalization to spatial symmetries, this analysis indicates that HOTIs remain well-defined on fractal lattices as long as symmetries protecting the parent state remain unbroken.

\section{Conclusions}
\label{sec:conclusions}

In this paper, we have presented general principles which determine the fate of 2D topological states on some fractal lattices, with numerics supporting our analytic arguments. Our results strongly suggest that lattices such as the SG (SC) can support gapless (gapped) topological phases, whose properties derive from those of an underlying parent state. Understanding the role of interactions remains an important open question, as does extending these ideas to 3D topological phases on \textit{e.g.}, the Sierpinski prism, where novel behavior could result from the rich structure of surface states. A more thorough investigation of the gapless topological phase of the $p+ip$ superconductor on the SG is also warranted and could shed light on its low-energy effective field theory as well as the observed \textit{topological} metal-to-insulator transition. Finally, given the progress in fabricating fractal lattices~\cite{herek,shang,tait2015,zhang2016,kempkes2018} and in realizing HOTIs on a variety of platforms~\cite{imhof2018,garcia2018,peterson2018,schindler2018}, experimentally realizing corner modes on a fractal lattice could be within reach.

\begin{acknowledgments}
We are especially grateful to Sheng-Jie Huang and Rahul Nandkishore for discussions which inspired this work. We also acknowledge stimulating conversations and correspondence with Yang-Zhi Chou, Victor Gurarie, Michael Hermele, Sergej Moroz, Titus Neupert, and Michael Pretko. The work of SP is supported by the U.S. Department of Energy, Office of Science, Basic Energy Sciences (BES) under Award number DE-SC0014415. AP is supported by a PCTS Fellowship at Princeton University.
\end{acknowledgments}


\appendix

\section{$p+ip$ on a Triangular Lattice}
\label{app:triangle}

While the $d=2$ BdG Hamiltonian describing the chiral $p + ip$ superconductor (Eq.~(1) in the main text) is typically implemented on a square lattice (see \textit{e.g.} Ref.~\onlinecite{tanaka2012}), it also allows for a topological phase on a triangular lattice, which we discuss briefly here. For a system with periodic boundary conditions along both $x$ and $y$ directions, we can write the Hamiltonian in momentum space as 
\beq
\hat{H} = \sum_{\bf{k}} 
\begin{pmatrix}
\hat{c}_{\bf{k}}^\dagger & \hat{c}_{-\bf{k}}
\end{pmatrix} \mathcal{H}_{\bf{k}}
\begin{pmatrix}
\hat{c}_{\bf{k}} \\
\hat{c}_{-\bf{k}}^\dagger
\end{pmatrix} \, ,
\eeq
where $\hat{c}_{\bf{k}}^\dagger$ and $\hat{c}_{\bf{k}}$ are fermionic creation and annihilation operators corresponding to momentum $\bf{k}$, and where 
\beq
\mathcal{H}_{\bf{k}}=\dfrac{1}{2}
\begin{pmatrix} 
\epsilon_{\bf{k}} & \Delta_{1,\bf{k}} \\
\Delta_{2,\bf{k}} & -\epsilon_{\bf{k}} 
\end{pmatrix},
\label{bdg1}
\eeq
with
\begin{widetext}
\begin{align}
\epsilon_{\bf{k}} &=-2t\bigg[\text{cos}(k_{x})+\text{cos}\bigg(\dfrac{k_{x}}{2}+\dfrac{\sqrt{3}k_{y}}{2}\bigg)+  \text{cos}\bigg(\dfrac{k_{x}}{2}-\dfrac{\sqrt{3}k_{y}}{2}\bigg)\bigg]-\mu
\label{bdg2} , \\
 \Delta_{1,\bf{k}} &=-2i \Delta \bigg[\text{sin}(k_{x})+e^{i\pi/3}\text{sin}\bigg(\dfrac{k_{x}}{2}+\dfrac{\sqrt{3}k_{y}}{2}\bigg)+  e^{2i\pi/3}\text{sin}\bigg(-\dfrac{k_{x}}{2}+\dfrac{\sqrt{3}k_{y}}{2}\bigg)\bigg] 
 \label{bdg3} , \\
  \Delta_{2,\bf{k}} &=2i \Delta \bigg[\text{sin}(k_{x})+e^{-i\pi/3}\text{sin}\bigg(\dfrac{k_{x}}{2}+\dfrac{\sqrt{3}k_{y}}{2}\bigg)+ e^{-2i\pi/3}\text{sin}\bigg(-\dfrac{k_{x}}{2}+\dfrac{\sqrt{3}k_{y}}{2}\bigg)\bigg] \, .
  \label{bdg4}
\end{align}
\end{widetext}

\begin{figure}[b]
 \centering
 \includegraphics[scale=0.3]{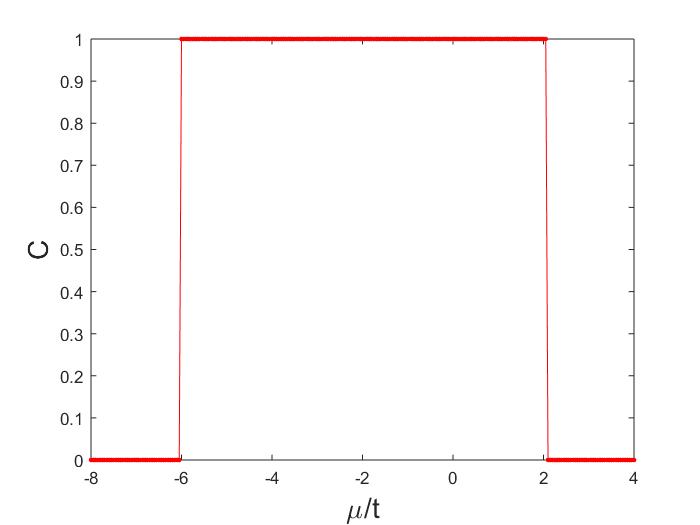}
 \caption{Chern number $C$ for the $p+ip$ superconductor on a triangular lattice. The plot shows that $C=1$ for $-6<\mu/t<2$ (the topological phase), and $C=0$ otherwise, \textit{i.e.} in the trivial phase. The above holds for any $\Delta \neq 0$.}
 \label{fig:triangleOBC}
 \end{figure}

The energy eigenvalues of $\mathcal{H}_{\bf{k}}$ are given by $E(\bf{k}) = \pm \sqrt{\epsilon_{\bf{k}}^{2}+\Delta_{1,\bf{k}}\Delta_{2,\bf{k}}}$. Here, $t$ is the hopping parameter, $\Delta$ is the pairing amplitude, and $\mu$ is the chemical potential. It is straightforward to check that this system has gap closings at $\mu = -6t, 2t$. For a triangular lattice with open boundary conditions, the above Hamiltonian gives rise to persistent chiral Majorana edge modes for $-6t < \mu < 2t$ for any $\Delta \neq 0$. These parameter values, therefore, characterize the trivial $\leftrightarrow$ topological transition on the triangular lattice. 

Writing the BdG Hamiltonian in Eq.~\eqref{bdg1} as $\mathcal{H}_{\bf{k}}=\bf{h}(\bf{k}) \cdot \bm{\sigma}$, with $\bf{h}(\bf{k})$ being a smooth function which is nonzero for all momenta, such that the bulk is fully gapped, we can then define a unit vector $\hat{\bf{h}}(\bf{k})$ that maps the 2D momentum space (defined on $\mathcal{T}^2$) onto a unit sphere. Here, $\bm{\sigma}$ is the usual vector of Pauli matrices $\sigma_i, \, i = x,y,z$. The momentum-space Chern number $C$ is then given by~\cite{volovik1988}
\begin{equation}
    C = \int_{\bf{k} \in \text{BZ}} \dfrac{d^{2}\bf{k}}{4\pi} \Big [ \hat{\bf{h}} \cdot \left( \partial_{k_{x}}\bf{\hat{h}}\times  \partial_{k_{y}}\bf{\hat{h}} \right) \Big] \, ,
\end{equation}
where ``BZ'' refers to Brillouin Zone. We find that $C=1$ in the topological phase ($-6t < \mu < 2t$), and $C=0$ in the trivial phase (see Fig.~\ref{fig:triangleOBC}).

\begin{figure}[t]
 \centering
 \includegraphics[scale=0.4]{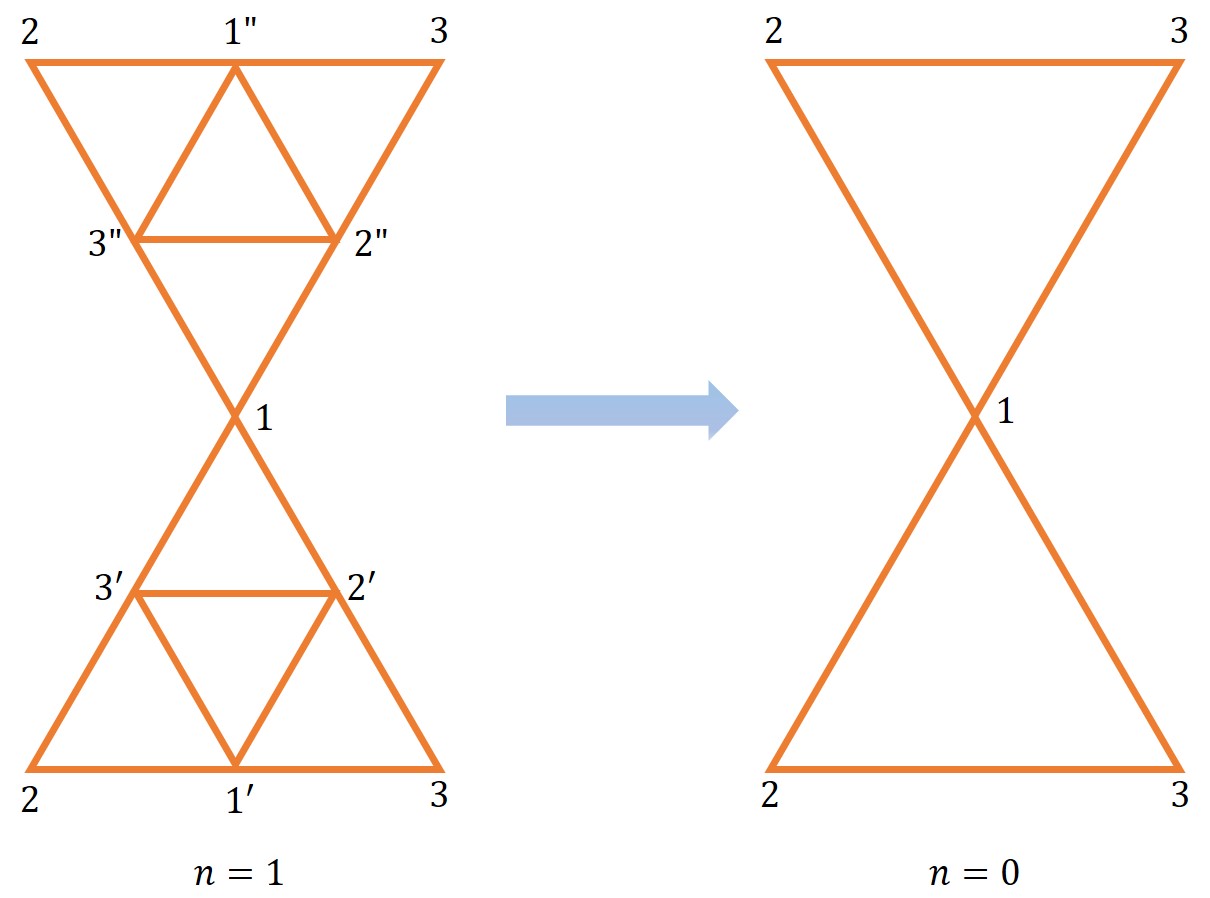}
 \caption{Decimating a $g=1$ lattice to a $g=0$ lattice by using Eq.~\eqref{effectiveH}.}
 \label{fig:recursionFig}
 \end{figure}

\begin{figure*}[t]
\centering
\includegraphics[scale=0.35]{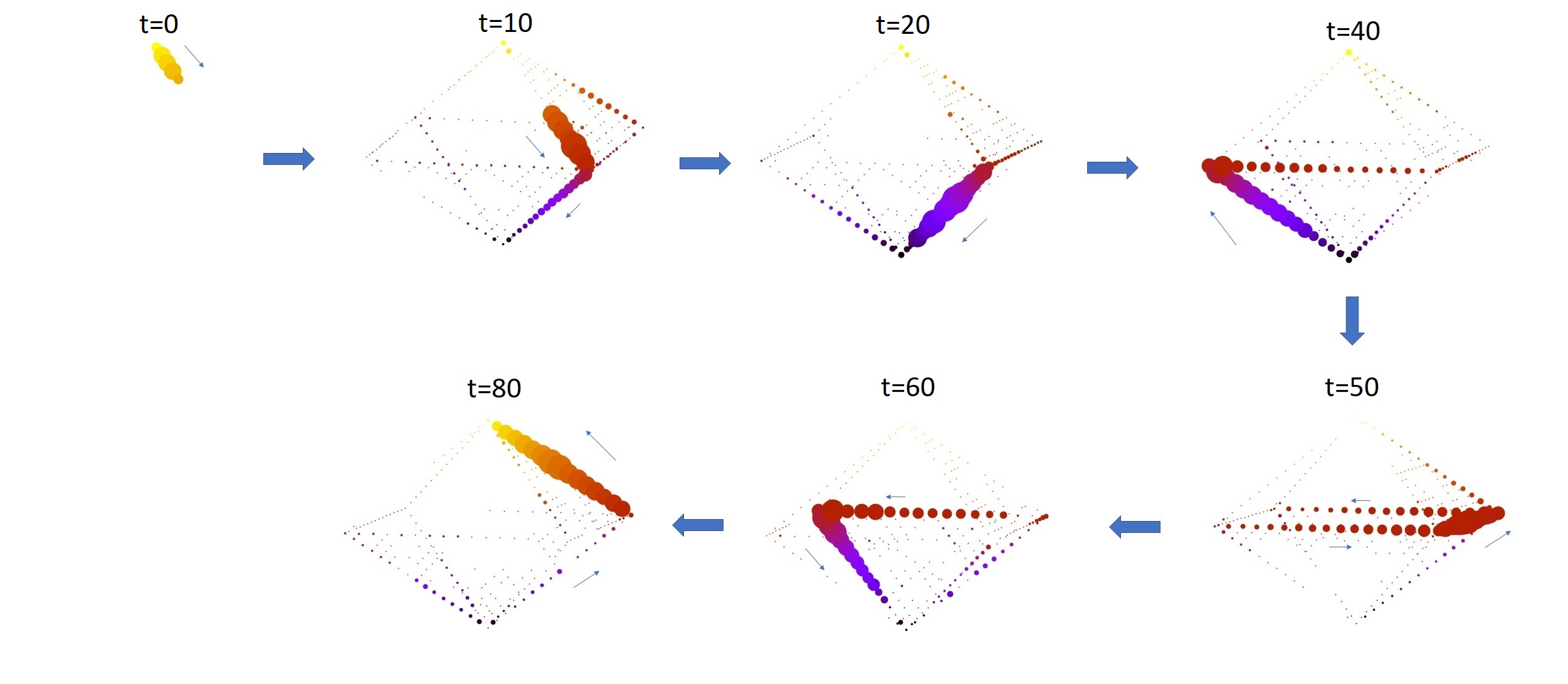}
 \caption{The evolution of a wave packet created by projecting onto edge states close to zero energy within the energy range ($-0.3 < \varepsilon < 0.3$) is shown. It can be seen that it moves exclusively on the edge of the system with definite chirality ($g=4, \mu=0.5, t=0.5, \Delta=0.5$).}
 \label{fig:chiral}
 \end{figure*}
 
\section{Recursive method for determining the BdG eigenspectrum on the Sierpinski Gasket}
\label{app:recursive}
 
We follow the analysis in Ref.~[\onlinecite{recursion}] to show that it is sufficient to study an effective model defined on a subset of the original sites rather than solving an eigenvalue equation involving all sites of the fractal lattice \textit{i.e.,} the SG. The eigenvalue equation for our system takes the form $H|\psi \rangle = E|\psi \rangle$. We divide the Hilbert space into two subspaces: one subspace consisting of all sites added \textit{up to} the $(n$-$1)^{th}$ generation, and the other subspace with sites added \textit{at} the $n^{th}$ generation. We refer to these subspaces as $A$ and $B$ respectively. We denote the projection of $|\psi \rangle$ onto these two subspaces as $|\psi_{A} \rangle$ and $|\psi_{B} \rangle$, with the eigenvalue equation then given by:
\begin{equation}
    \begin{pmatrix} 
H_{AA} & H_{AB} \\
H_{BA} & H_{BB} 
\end{pmatrix}
\begin{pmatrix} 
|\psi_{A} \rangle \\
|\psi_{B} \rangle 
\end{pmatrix} = E \begin{pmatrix} 
|\psi_{A} \rangle \\
|\psi_{B} \rangle 
\end{pmatrix} \, ,
\end{equation}
following which we can can formally write
\begin{equation}
 |\psi_{B} \rangle = (E-H_{BB})^{-1}H_{BA}|\psi_{A} \rangle \, .
\end{equation}
As discussed in Ref.~[\onlinecite{recursion}], we can now define an ``effective'' Hamiltonian acting only on the sites of the decimated lattice \textit{i.e.,} sites belonging to subspace $A$:
\begin{equation}
\label{effectiveH}
    H^{\textrm{eff}}|\psi_{A} \rangle = [H_{AA} + H_{AB} (E-H_{BB})^{-1}H_{BA}]|\psi_{A} \rangle \, .
\end{equation}

We now apply this formalism to the system under consideration. An additional feature of the BdG Hamiltonian in Eq.~\eqref{bdg1} is the presence of two ``orbitals'' per site instead of one. To obtain the analogue of Eq.~\eqref{effectiveH}, we need the hopping matrices associated with the underlying $n=1$ triangle (see Fig.~\ref{fig:recursionFig}).

For $i,j \in \{1,2,3\}$, we find that
\begin{equation}
\begin{split}
    (H_{AA})_{ij} &= -\dfrac{\mu}{2}\sigma_{z}\delta_{ij} \, ,\\ 
    (H_{AB})_{ij} &= -\dfrac{\mu}{2}\sigma_{z}\delta_{ij} + (1-\delta_{ij})\bigg[-\dfrac{t}{2}\sigma_{z}-i\Delta e^{i\alpha_{ij}}\sigma_{y}\bigg] \\ &= (H_{BA})_{ij} = (H_{BB})_{ij} \, ,
\end{split}
\end{equation}
for the Hamiltonian defined in Eqs.~\eqref{bdg1}-\eqref{bdg4}. Here, $\alpha_{ij}$ is the angle between the link joining sites $i$ and $j$ and the local $x$-axis at site $i$. Using Eq.~\eqref{effectiveH}, we find that
\begin{equation}
     H^{\textrm{eff}}_{ij} = [H_{AA} + H_{AB} (E-H_{BB})^{-1}H_{BA}]_{ij} \, .
     \label{effectiveH2}
\end{equation}

Since the BdG Hamiltonian gives rise to robust chiral edge states for $-6t < \mu< 2t$ for any $\Delta \neq 0$, we set $\Delta=1$ and $\mu=0$ here (corresponding to the topological phase for any nonzero $t$) to simplify our analysis. Other parameters can be analyzed following the procedure delineated here. Now, we compare the \textit{effective} Hamiltonian with the original BdG Hamiltonian but now defined on the generation $n=0$ lattice and with hopping parameter $t'$. This allows us to express the effective Hamiltonian as the BdG Hamiltonian acting on sites in the $A$ sublattice, but with renormalized hopping strength $t'$. Using Eq.~\eqref{effectiveH2}, we can derive an expression for $t'$ in terms of the original parameters:

\begin{widetext}
\begin{equation}
t' = \dfrac{t(48-12t^{2}+t^{6})-t(144+7t^{2}(4+t^{2}))E^{2}+2t^{2}(-10+t^{2})E^{3}+4t(16+3t^{2})E^{4}-8(2+t^{2})E^{5}}{48-12t^{2}+t^{6}-3(48+8t^{2}+3t^{4})E^{3}+4t(16+3t^{2})E^{4}-16E^{6}} \, 
    \label{teff}
\end{equation}
\end{widetext}

Next, we use Eq.~\eqref{teff} and the relation $ t' \, \epsilon_{n-1} = t \,\epsilon_{n}$ to derive a recursion relation between $\epsilon_{n-1}$ and $\epsilon_{n}(=E/t)$, the dimensionless (scaled by the hopping energies $t'$ and $t$ respectively) energy eigenvalues on the generation $n-1$ and $n$ SGs. Therefore, in principle, given an energy eigenvalue $\epsilon_{n-1}$ of the system defined on the generation $(n-1)$ SG, Eq.~\eqref{teff} allows us to determine the corresponding eigenvalues on the generation $n$ lattice. However, as pointed out in Ref.~[\onlinecite{recursion}], the recursion relation by itself does not give the correct degeneracy for those eigenvalues which correspond to the zeroes of the denominator: these have to be put in by hand at every iteration of the recurrence relation.

\section{Chiral nature of eigenstates in the topological phase}
\label{app:chiral}

In order to visualize the chiral nature of the edge modes that appear in the topological phase of the $p+ip$ superconductor on the SG, we construct an initial wave packet localized over a few sites belonging to some outer edge of the Sierpinski gasket, and project it onto edge states within an arbitrary but small energy window close to zero, say $(-0.3< \varepsilon < 0.3)$, in order to obtain the propagating edge mode shown in Fig.~\ref{fig:chiral}. 
 
\begin{figure*}[t]
\centering
\includegraphics[scale=0.52]{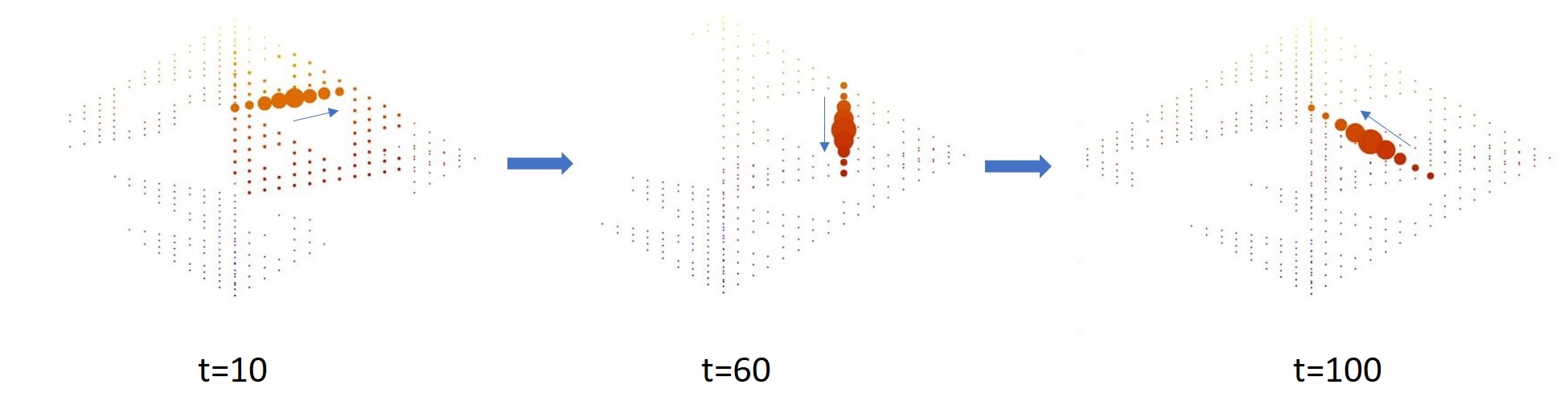}
 \caption{The evolution of a wave packet initialized on an inner edge is shown. It can be seen that it moves exclusively on the corresponding inner edge of the system with chirality opposite to that of the outermost edge ($g=4, \mu=0.5, t=0.5, \Delta=0.5$).}
 \label{fig:chiralinner}
 \end{figure*}

Likewise, we project a wave packet localized on an inner edge onto the states within a similar energy range in one of the other gaps in the spectrum to obtain Fig.~\ref{fig:chiralinner}. We find that the chirality of wave packets initialized on any of the inner edges (or holes) of the lattice is opposite to that of a wave packet propagating along the outermost edge. 

\section{Scaling of the gap in the topological phase}
\label{gap:scaling}

\begin{figure}[b]
 \centering
 \includegraphics[scale=0.6]{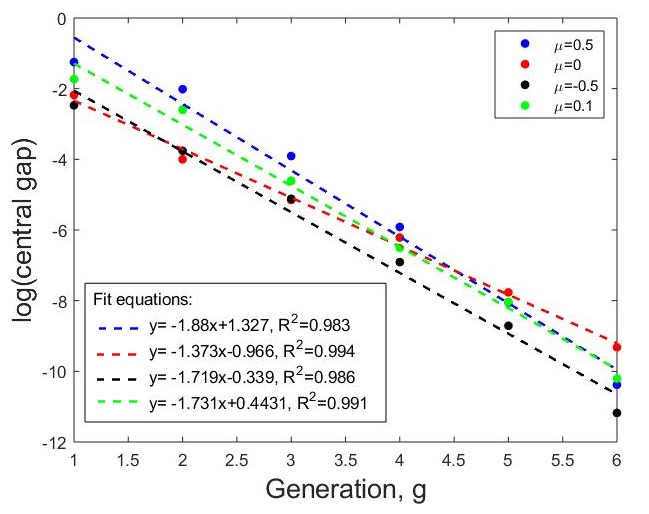}
 \caption{Scaling of the gap as a function of the generation $g$ of the Sierpinski gasket in the topological phase. The gap decays to zero exponentially fast as $g \to \infty$. Here, $t=0.5, \Delta=0.5$.}
 \label{fig:gapless}
 \end{figure}

For any finite generation $g$, the spectrum of the $p+ip$ state on the SG presents a finite number of gaps $\{\mathcal{E}_j\}$. However, the amplitude of these gaps decreases \textit{exponentially} with $g$, such that the spectrum becomes gapless in the thermodynamic limit. Specifically, we have the analyzed the scaling of the largest gap in the spectrum as a function of $g$, for various parameters corresponding to the topological phase. In Fig. \ref{fig:gapless}, we show the gap scaling on a semi-log plot, which clearly demonstrates that the largest gap in the spectrum goes to zero exponentially fast as $g \to \infty$ \textit{i.e.,}
\beq
\textrm{max}_j \mathcal{E}_j = \mathcal{E}_{\textrm{max}} \sim \Delta e^{-\beta g} \, ,
\eeq
for some $\beta >0$, which is weakly dependent on $\mu/t$. Since the maximal gap $\mathcal{E}_{\textrm{max}} \to 0$ as $g\to \infty$, all the other gaps also vanish, leading to a gapless phase in the thermodynamic limit.

\section{Details regarding the decimation procedure}
\label{app:decimation}

\subsection{SG from triangular lattice}

As discussed in the main text, we consider a triangular lattice with lattice spacing $a$. We assume that the lattice takes the shape of an equilateral triangle with each side of length $L = 2^p a$ ($p \in \mathbb{Z}^+$). The thermodynamic limit is taken in the usual way, $L/a \to \infty$. The coordination number of sites in the interior of the lattice, which we denote bulk sites, equals six, while that of those along the edge, denoted edge sites, equals four. To ensure that the three corner sites, which have coordination number two, are also boundary sites, we can place four copies of this lattice in the arrangement depicted in Fig.~1(a) of the main text. For simplicity, we discuss the recursive decimation for a single lattice here, with the analysis carrying over as is for that configuration. Alternatively, we can also simply count the corner sites as boundary sites; since the ratio of corner sites to boundary sites vanishes in the thermodynamic limit, this will not affect our analysis. Defining $l\equiv L/a$, the number of boundary sites is hence $3l$ while the number of bulk sites is $\dfrac{1}{2}(l-2)(l-1)$, with only a single, outer boundary present. 

At the first step of the decimation procedure, we eliminate sites and bonds contained in the interior of a inverted triangular lattice with side $L/2$, which introduces an interior boundary into the lattice. At the $g^{th}$ step ($g\geq 1$), we eliminate all sites and bonds contained within $3^{g-1}$ inverted lattices of length $L/2^g$, which are arranged self-similarly within the parent triangular lattice (see Fig.~5 in the main text). This introduces an additional $3^{g-1}$ boundaries into the parent lattice, such that the total number of boundaries at step $g$ is $\dfrac{1}{2}(3^g + 1)$, which includes the single outermost boundary of the underlying triangular lattice. Denoting the number of bulk and edge sites present at the $g^{th}$ iteration as $n_B(g)$ and $n_E(g)$ respectively, straightforward algebra shows that
\begin{align}
    n_B(g) = & \, \frac{1}{2}(l-2)(l-1) - \frac{1}{2} \sum_{j=1}^g 3^{j-1} \left(\frac{l}{2^j} -2 \right)\left(\frac{l}{2^j} -1 \right) \nonumber \\ 
    & \, - \sum_{j=1}^g 3^{j-1} \left(\frac{3 l}{2^j} - 3 \right) \, , \\
    n_E(g) = & \, 3l + \sum_{j=1}^g 3^{j-1} \left(\frac{3 l}{2^j} - 3 \right) \, .
\end{align}

The SG is generated once \textit{all} sites are edge sites with coordination number four. Hence, we stop the process once we have eliminated the smallest triangle containing sites and bonds contained within its interior \textit{i.e.,} at step $g = g_c$, with $2^{g_c} = \frac{l}{2},$ since a triangle with side length $a$ is the smallest possible triangle and contains no interior sites or bonds. It is then easy to check that $n_B{(g_c)} = 0$ as stated in the main text. In the thermodynamic limit, $g_c \to \infty$ so that the decimation process must be repeated infinitely many times, leading to infinitely many inner edges created within the parent triangular lattice. Moreover, for a fixed $l = 2^p$, one can check that 
\beq
\lim_{g\to p}\frac{n_B(g)}{n_E(g)} = 0 \, .
\eeq

\subsection{SC from square lattice}
\label{app:carpet}

\begin{figure}[b]
    \centering
    \includegraphics[width=0.45\textwidth]{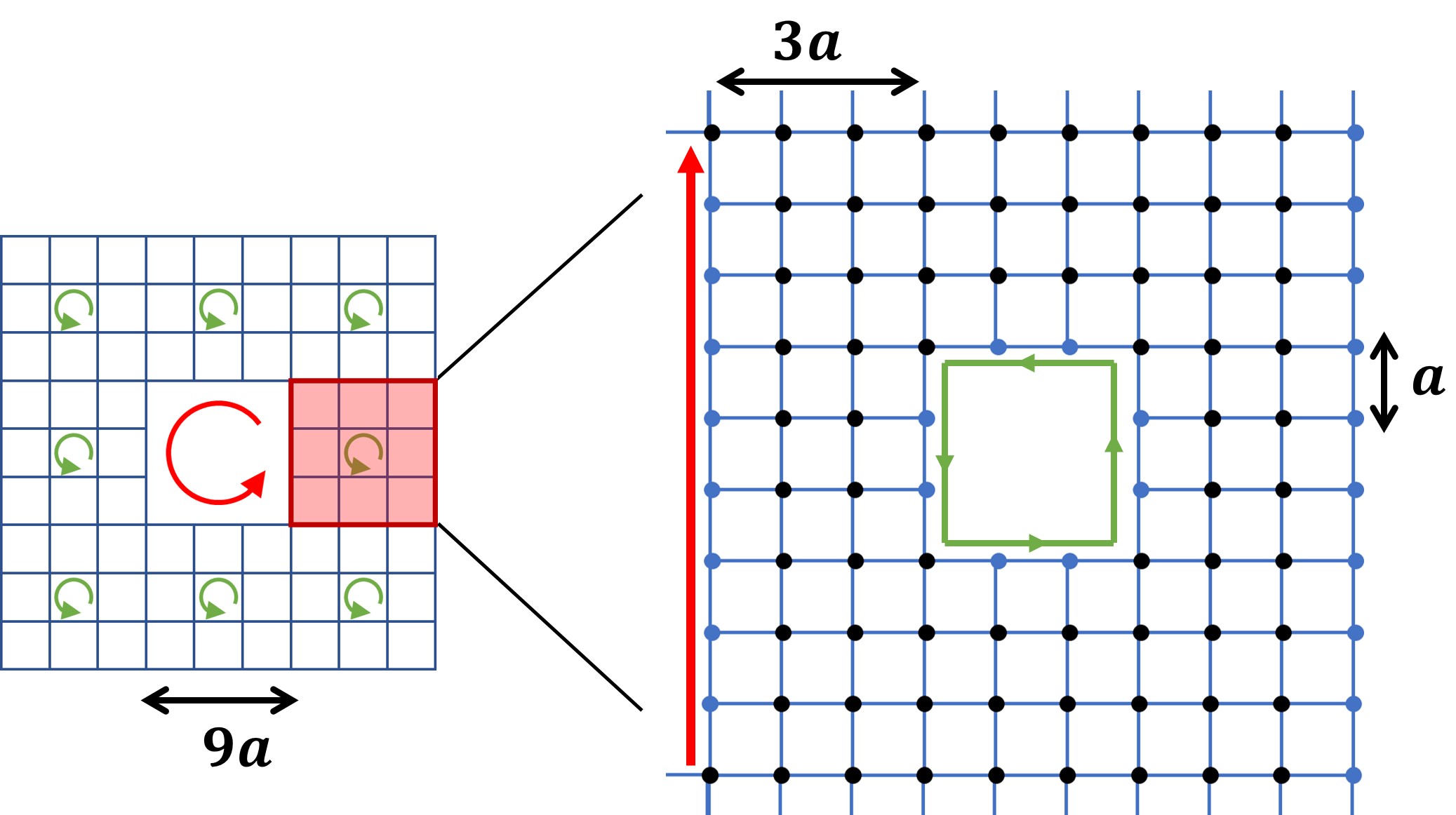}
    \caption{Generating an SC from a square lattice. The first (second) step results in inner edge mode(s), shown in red (green), when starting from the topological phase of the $p+ip$ superconductor on a square lattice. However, a finite number of bulk sites (black dots) remain even after the SC is generated, as shown in the zoomed in image on the right.}
    \label{fig:carpet}
\end{figure}

\begin{figure*}[t]
\centering
\begin{subfigure}[t]{0.3\textwidth}
\includegraphics[width=\textwidth]{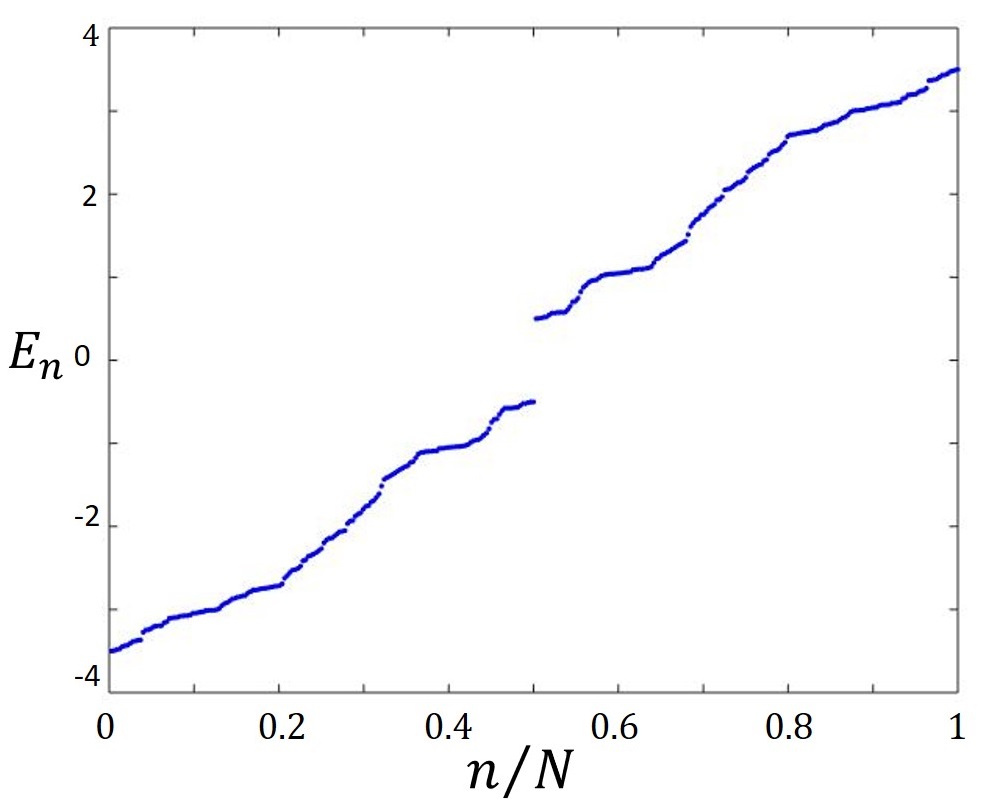}
\caption{}
\label{fig:carpetTriv}
\end{subfigure}\quad
\begin{subfigure}[t]{0.3\textwidth}
\includegraphics[width=\textwidth]{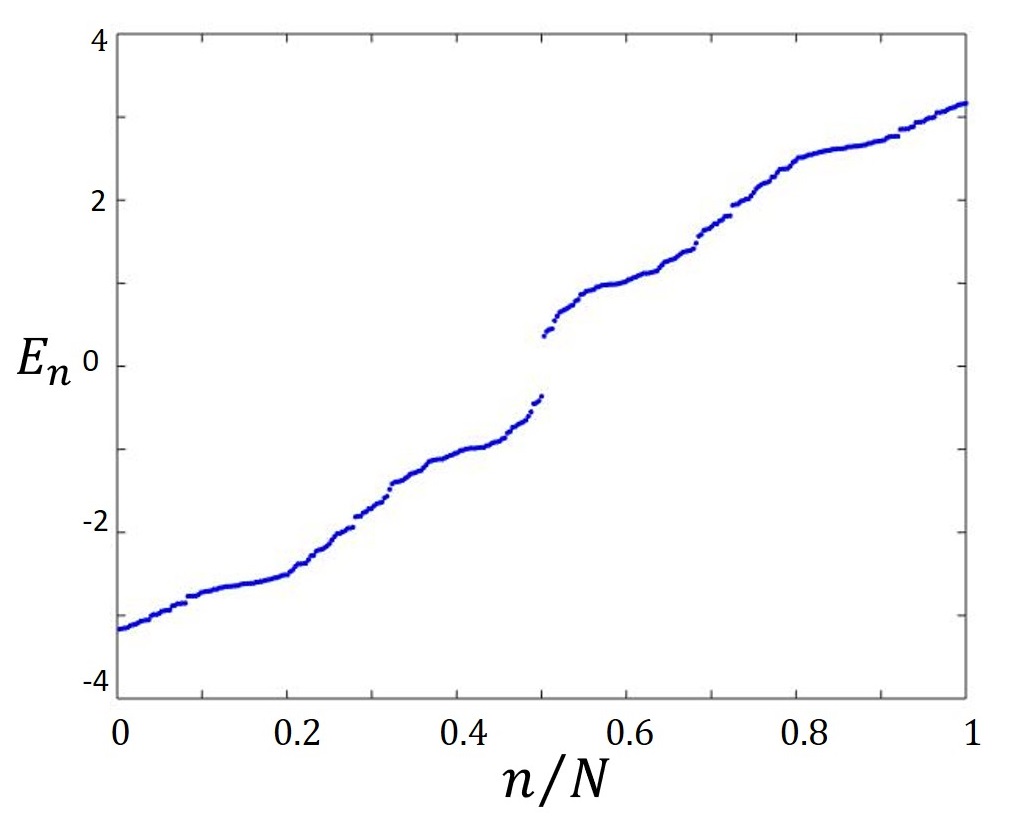}
\caption{}
\label{fig:carpetTop}
\end{subfigure}\quad
\begin{subfigure}[t]{0.3\textwidth}
\includegraphics[width=\textwidth]{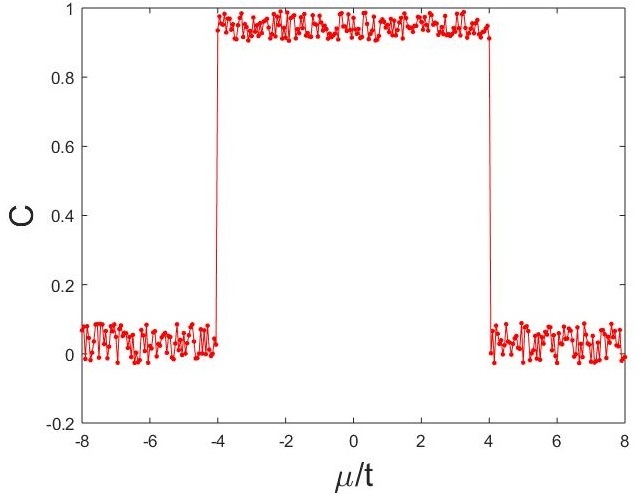}
\caption{}
\label{fig:carpetChern}
\end{subfigure}
\caption{$p+ip$ state on the Sierpinski Carpet: the energy eigenvalue spectrum is shown for (a) the trivial phase $(t=0.5, \mu=1,)$, and (b) the topological phase $(t=0.5, \mu=0.25)$. (c) shows the real space Chern number within the gap as function of $\mu/t$, clearly indicating the existence of a gapped topological phase on the SC.}
\end{figure*}

We now repeat the above analysis in order to generate a lattice-regulated SC (with a smallest square) from a square lattice through recursive decimation. A key distinction between the SG and the SC is that the former has a \textit{finite} ramification while the latter is \textit{infinitely} ramified; in other words, only a finite number of bonds need to be cut to separate out an extensive piece of the gasket, while for the carpet, the number of bonds which need cutting tends to infinity in the thermodynamic limit. Crucially, while the SG at any generation has only edge sites, the SC always contains a finite number of sites with bulk coordination number. The procedure follows that discussed in the previous section closely: consider a square lattice with lattice spacing $a$ and length $L$, with $l = L/a$. Bulk and edge sites have coordination number four and three respectively, where we again subsume corner sites with coordination number two as boundary sites since the ratio of corner sites to edge sites vanishes as $l \to \infty$. The parent lattice thus has $4l$ boundary sites and $(l-2)^2$ bulk sites, with a single outer boundary. 

At the $g^{th}$ step ($g \geq 1$) of the decimation, we eliminate sites and bonds contained within $8^{g-1}$ square lattices of length $L/3^g$, arranged self-similarly within the parent square lattice (see Fig.~\ref{fig:carpet}). This introduces an additional $8^{g-1}$ inner edges into the parent lattice, such that the total number of boundaries at step $g$ is $\dfrac{1}{7}(8^g + 6)$, including the outermost boundary of the underlying square lattice. As before, denoting the number of bulk and edge sites present at the $g^{th}$ iteration as $n_B(g)$ and $n_E(g)$ respectively, we find that 
\begin{align}
    n_B(g) =& (l-2)^2 - \sum_{j=1}^g 8^{j-1} \left(\frac{l}{3^g} -1 \right)^2  \nonumber \\
    &- \sum_{j=1}^g 8^{j-1} \left(\frac{4 l}{3^g} - 4 \right)\, , \\
    n_E(g) =& 4l + \sum_{j=1}^g 8^{j-1} \left(\frac{4 l}{3^g} - 4 \right)\, 
\end{align}
We arrive at the SC when $g = g_c$, with $3^{g_c} = \frac{l}{3}$. In the thermodynamic limit, we hence require $g_c \to \infty$, with 
\beq
\lim_{g_c \to \infty}\frac{n_B(g_c)}{n_E(g_c)} = \frac{315}{64} \sim 5,
\eeq
such that the ratio of bulk to edge sites remains finite.

Following the above analysis, it is also straightforward to see that at the $g^{th}$ step of the decimation procedure, each chiral edge mode is separated from another one by a distance $L/3^g$. At the final step $g = g_c$, where the SC is generated, each mode is separated by $3a$ from an edge mode with \textit{opposite} chirality, as illustrated in Fig.~\ref{fig:carpet}. Since the separation between such counter-propagating Majorana edge modes approaches their bulk penetration depth at large $g$, these states are gapped out due to back-scattering, resulting in a gapped spectrum. As discussed in the main text, the hybridization of the gapless edge states is a consequence of a non-vanishing ratio of bulk to boundary coordinated sites in the thermodynamic limit, which in turn allows the SC to host gapped topological phases retaining the bulk features of the phase defined on the parent square lattice.

\section{Numerical diagonalization of the BdG Hamiltonian on the SC}

The pairing term of the BdG Hamiltonian (Eq.~(1) in the main text) on a square lattice is specified by $\Delta_{\hat{x}} = \Delta$ and $\Delta_{\hat{y}} = i \Delta$, defined on the nearest-neighbor bonds corresponding to the lattice vectors $\bm{e}_{\hat{x}}$ and $\bm{e}_{\hat{y}}$. The spectrum is gapped everywhere for $\Delta \neq 0$, except at $\mu=\pm 4t$, with $|\mu|<4t$ corresponding to the topological phase, which has a quantized momentum space Chern number $C = 1$ and hosts a chiral gapless Majorana mode along the sample boundary. We numerically diagonalize this model on the Sierpinski carpet and find that, unlike the model on the SG, the spectrum remains gapped in both the trivial ($|\mu| > 4t$) and the topological phase ($|\mu|<4t$), as shown in Figs.~\ref{fig:carpetTriv} and~\ref{fig:carpetTop} respectively. We also calculate the real space Chern number (Eq.~(2) in the main text) within the gap as a function of $\mu/t$ and find that it vanishes in the trivial phase, but takes on a quantized value $\mathcal{C} = 1$ in the topological phase, as shown in Fig.~\ref{fig:carpetChern}.



\bibliography{library}

\end{document}